\documentclass[12pt]{book}
\usepackage{myplenumQPH}
\begin{document}

\newcommand{\be}{\begin{equation}}
\newcommand{\ee}{\end{equation}}

\chapter{JUST {\it TWO\/} NONORTHOGONAL QUANTUM STATES}

\author{Christopher A. Fuchs}

\affiliation{Norman Bridge Laboratory of Physics, 12-33\\
California Institute of Technology\\
Pasadena, California 91125\smallskip\\
email: cfuchs@cco.caltech.edu}
\bigskip\bigskip

\abstract{From the perspective of quantum information theory, a 
system so simple as one restricted to just two nonorthogonal states 
can be surprisingly rich in physics. In this paper, we explore the 
extent of this statement through a review of three topics: (1) 
``nonlocality without entanglement'' as exhibited in binary quantum 
communication channels, (2) the tradeoff between information  gain 
and state disturbance for two prescribed states, and (3) the 
quantitative clonability of those states.  Each topic in its own way 
quantifies the extent to which two states are ``quantum'' with 
respect to each other, i.e., the extent to which the two together
violate some classical precept.  It is suggested that even toy 
examples such as these hold some promise for shedding light on the 
foundations of quantum theory.}

\section{INTRODUCTION}

The total set of states available to a quantum system corresponds to
the uncountably infinite set of density operators over a given 
Hilbert space.  With that set and a sufficiently general 
notion of measurement and time evolution, one can say 
{\it everything\/} that can be said about the system.%
\refnote{\cite{Kraus83}}  \ Nevertheless, as one gains
experience in our field, it becomes hard to shake the feeling that 
much of the {\it essence\/} of quantum theory already makes itself
known in the case of just two nonorthogonal states.%
\refnote{\cite{Pitowsky98}} \ This is because of the overpowering 
importance of the quantum no-cloning theorem%
\refnote{\cite{Wootters82}}: a set of two 
nonorthogonal states is the smallest set of states for which the
theorem is active.\refnote{\cite{Yuen86}} \ 
More generally, such a set forms the smallest set of states for which
{\it no\/} information can be gathered without a conjugate 
disturbance.\refnote{\cite{Bennett92b}} \ They fulfill a role 
that the founding fathers tried so hard to pin on a single, solitary
quantum state.\refnote{\cite{Fuchs98}}

In this connection, the ultimate questions we should like to ask 
are the following.  To what extent does the newfound 
language of quantum information allow us to sharpen our understanding
of this example and, more importantly, what can it tell us about the 
foundations of quantum theory itself?  What hint might it give us of 
the tools required for digging even deeper in the coming century?
These, of course, are difficult questions.  But certainly no progress
can be made in their answering without the courage of one small step.
Here, we shall start in that direction by reviewing what is known 
about two nonorthogonal states that is
already expressible in the language of quantum information.  
In particular, we will pay attention to how this allows us to 
express when two states are the most ``quantum'' with respect to 
each other.  We will do this, in turn, from the perspective of
(1) ``nonlocality without entanglement'' in binary quantum 
channels, (2) the tradeoff between information gain and state 
disturbance in quantum eavesdropping, and (3) the imperfect
clonability of two states by various criteria.  At the paper's 
conclusion, we will use these perspectives to attempt a tighter 
formulation of the grand questions above.

Throughout we will consider two nonorthogonal pure states 
$|\psi_0\rangle$ and $|\psi_1\rangle$ separated in a Hilbert space
${\cal H}_S$ by some angle $\theta$.  Without loss of generality for
our considerations, we assume that the overlap
$
x=\langle\psi_0|\psi_1\rangle=\cos\theta
$
is a positive real number.  The variable $x$ will be the most 
important parameter for our problems, expressing in one way or the 
other the degree of quantumness the two states hold with respect to 
each other.  For the problems below that require the assumption of
some a priori probabilities for the two states, we 
will assume them equal.  To say that the identity of a state
is ``unknown'' is to say that $|\psi_0\rangle$ and 
$|\psi_1\rangle$ are each assigned an a priori probability 
of 1/2\@.  Whenever it is more convenient to think of the
two quantum states as density operators, we shall denote them by
$\rho_0=|\psi_0\rangle\langle\psi_0|$ and 
$\rho_1=|\psi_1\rangle\langle\psi_1|$. 

\section{NONLOCALITY w/o ENTANGLEMENT FOR BINARY CHANNELS}

Consider using the alternate preparations $|\psi_0\rangle$ and 
$|\psi_1\rangle$ as the physical implementation of a binary alphabet
in some communication scheme.  Why would one ever want to do this?
Well, there are various reasons based on practical considerations.
For instance, the transmitter may have only low-energy coherent 
states available for signaling---these are necessarily 
nonorthogonal.\refnote{\cite{Sasaki96}} \ Also, nonorthogonal signals
are sometimes able to achieve higher capacities in noisy quantum 
channels than orthogonal signals.%
\refnote{\cite{Fuchs97}} \ But let us consider this
possibility purely for its own aesthetics.

With the adoption of a nonorthogonal alphabet, the signals will,
of necessity, be imperfectly distinguishable by the receiver.  
For instance, 
if the criterion of distinguishability is the smallest possible 
error probability $P_e$ in a decision about the signal's 
identity (following some measurement),
then\refnote{\cite{Fuchs98},\cite{Helstrom76}}
\be
P_e \,=\, \frac{1}{2}- \frac{1}{4}{\rm tr}|\rho_1-\rho_0|
\,=\,\frac{1}{2}\Big(1-\sqrt{1-x^2}\Big)\;.
\label{BettyBoop}
\ee
This measure, in fact, shows just what one expects: as the 
overlap  between the states increases, their mutual 
distinguishability decreases.

What is {\it quantum\/} about this lack of distinguishability in the
signal states?  One might be tempted to say, ``Everything.''  If one 
draws an analogy between the quantum state and a point in a classical
phase space, then one has that classical states can always be 
distinguished with perfect reliability and quantum states cannot.  
However, a more proper analogy is between quantum states and 
the Liouvillean probability distributions on phase space.%
\refnote{\cite{Caves96}} \
That is to say, overlapping quantum states are more analogous to the
outputs of a noisy classical communication channel, where the 
receiver must distinguish between two probability distributions 
$p_0(y)$ and $p_1(y)$ over the output alphabet.  From this point of 
view, the answer to the question above is, ``Nothing.''  The 
distinction between quantum and classical must be seen in other
ways.

One natural way crops up in a different aspect of the communication 
scenario: it
is in the concept of {\it nonlocality without entanglement}.%
\refnote{\cite{Bennett98}} \ As the signals in a long message start
to accumulate, the receiver may be tempted to start the decoding
process signal by signal.  For classical channels, where the
task is to accumulate information about long strings of the
probability distributions $p_0(y)$ and $p_1(y)$, it turns out that 
such a strategy is never harmful.  Signal-by-signal decoding never 
decreases the number of reliable bits per 
transmission.\refnote{\cite{Cover91}} \ For quantum mechanical 
messages composed from a nonorthogonal alphabet, however, this is not
the case.  A higher channel capacity can be achieved by allowing the 
receiver the capability to perform large collective quantum 
measurements on multiple transmissions.\refnote{\cite{Holevo79}} \

More specifically, if the receiver is restricted to perform a fixed 
generalized measurement signal by signal, or even an 
adaptive measurement signal by signal,\refnote{\cite{Fujiwara98}}  
the greatest capacity achievable with a fixed alphabet is
given numerically by the alphabet's {\it accessible information\/}%
\refnote{\cite{Levitin95}} maximized over all 
a priori probability distributions.  In our case, this number turns 
out to be
\be
C_1=\frac{1}{2}\Big(1+\sqrt{1-x^2}\Big)\log\Big(1+
\sqrt{1-x^2}\Big)+\frac{1}{2}\Big(1-\sqrt{1-x^2}\Big)\log\Big(1-
\sqrt{1-x^2}\Big)\;.
\ee
On the other hand, if the receiver can perform collective quantum 
measurements over arbitrarily large numbers of signals, then the 
greatest capacity is calculated by maximizing the alphabet's von 
Neumann entropy over all a priori probability distributions.%
\refnote{\cite{Hausladen96}} \ The resultant in our case is%
\refnote{\cite{Fuchs96a}}
\be
C_\infty=1-\frac{1}{2}\Big((1-x)\log(1-x)+(1+x)\log(1+x)\Big)
\;.
\ee

The meaning of this result is that when one is speaking of
correlations between nonorthogonal states---as one would be in
the situation where these states are concatenated into codewords
for a communication channel---the whole is greater than the sum of 
the parts.  Extra correlation, and hence extra information, can be 
ferreted out of these codewords by collective measurements on the 
whole.%
\footnote{It is an open question whether these channel
examples exhibit the strongest form of ``nonlocality without
entanglement.''  In the strongest version,%
\refnote{\cite{Bennett98},\cite{Peres91}} one is not only concerned
with the discrepancy between collective and sequentially adaptive 
measurements, but between collective measurements and any 
measurements whatsoever that are purely local with respect to the 
separate transmissions.  For instance, within the largest class of
local measurements the  receiver might perform weak measurements that
ping-pong back and forth between the separate transmissions:  first 
collect a little information from signal 1, then adjust the 
measurement and collect a little information from signal 2.  Following
that, adjust again and return to signal 1 to collect a little more, 
and so on and so on.}  When the signals are orthogonal to each 
other---a situation in which one is tempted to say that they are 
classical---then the whole possesses nothing that the parts do not 
already contain.

This distinction in channel capacities suggests that the difference
\be
Q=C_\infty-C_1
\ee
defines an effective notion of ``quantumness'' for the two states.
It signifies the extra information the two states carry with respect
to each other that can be unlocked only by nonlocal means on separate
transmissions.

Notice that, by this reckoning, two states are the most quantum with 
respect to each other when $x=1/\sqrt{2}\,$, i.e., when the two 
states are separated by an angle $\theta=45^\circ$.  Here
$Q\approx0.202$.  In ways, this result is quite pleasing.  Since 
$C_\infty=C_1$ when $\theta=0^\circ$ and  $\theta=90^\circ$, one 
might well expect the states to be maximally 
quantal when their separation is exactly between these two extremes
in the sense of Hilbert-space geometry.

\section{INFORMATION GAIN vs.\ QUANTUM STATE DISTURBANCE}

The founding fathers of quantum mechanics were fond of saying things
like this typical example of Pauli's:%
\refnote{\cite{Pauli95}} 
\begin{quotation}
\small
\baselineskip=12pt
\noindent
The indivisibility of elementary quantum processes ...\ finds 
expression in an indeterminacy of the interaction between [the] 
instrument of observation ...\ and the system observed ..., which 
cannot be got rid of by determinable corrections.  For every act 
of observation is an interference, of undeterminable extent ...
\end{quotation}
However, given the difficulty in ascribing objective properties to 
quantum systems independently of measurement (as indicated by the 
Kochen-Specker theorem and the violation of the Bell inequalities%
\refnote{\cite{Peres93}}), what can the terminology of 
``interference'' or ``disturbance'' possibly refer to?  What 
precisely is it, if anything, that is disturbed by measurement?

One of the more interesting things about the applied field of quantum
cryptography as far as the foundations of quantum mechanics is
concerned is that it provides the tools to breath some real life 
into this old question.  To get somewhere, however, one must
realize that one cannot simply speak of performing measurements on a 
single quantum system prepared in a single quantum state: one must, 
at the very least, consider two nonorthogonal states as in the B92 
protocol.\refnote{\cite{Bennett92a}} \ Then the referents of the
words ``information gain'' and ``disturbance'' can have precise
meanings.

The scheme is the following.  Alice encodes the various secret bits 
she wishes to share with Bob into the quantum states $|\psi_0\rangle$
and $|\psi_1\rangle$ and sends them on their way.  The 
eavesdropper Eve interacts some probe with the systems while they
are en route.  This leaves her probe variously either in 
one of two (mixed) quantum states, 
$\rho_0^{\rm\scriptscriptstyle E}$ and 
$\rho_1^{\rm\scriptscriptstyle E}$.  In the process, Alice's states
are perturbed variously into $\rho_0^{\rm\scriptscriptstyle A}$ and 
$\rho_1^{\rm\scriptscriptstyle A}$.  

These four states taken together provide a basis for an 
information-disturbance tradeoff principle.  For instance, one might
gauge the amount of information that Eve has received by her
potential for guessing the individual key bits through measurements
on her probe.  Her best probability of error $P$ will given by the
leftmost expression in Eq.~(\ref{BettyBoop}) with the density 
operators suitably replaced by $\rho_0^{\rm\scriptscriptstyle E}$ and
$\rho_1^{\rm\scriptscriptstyle E}$.  Similarly one might gauge the
disturbance $D$ to Alice's system by Bob's probability of identifying 
Eve's intervention as he performs the standard maneuvers for
extracting a key from Alice's signals.  Holding $P$ fixed while
optimizing Eve's probe's interaction, one obtains the 
rather complicated tradeoff principle:\refnote{\cite{Fuchs98}}
\be
D=\frac{1}{2}-\frac{1}{2}\sqrt{1+x^2\!\left(-1-4P+4P^2+2x^2+
2\sqrt{(1-x^2)(4P-4P^2-x^2)}\right)}\;.
\ee

At the endpoint corresponding to a maximal information gain by Eve,
this tradeoff is especially interesting for defining a notion
of ``quantumness'' .  There, Eve's probability of error in
identifying the bit must be given by Eq.~(\ref{BettyBoop});
the minimal disturbance that can be imparted to Alice's
states in this case is
\be
D_{\rm\scriptscriptstyle @MI}=\frac{1}{2}\Big(1-\sqrt{1-x^2+x^4}\Big)\;.
\ee
This number---the minimum disturbance at maximum 
information---expresses the two states' relative fragility 
when exposed to information-gathering measurements.  Notice that
once again the two states are most quantum with respect to each other
when $x=1/\sqrt{2}\,$.  In that case, 
$D\approx0.067$ while Eve's probability of error is
$P_e\approx0.146$.  The angle $\theta=45^\circ$ 
starts to look quite robust as far as ``most quantum'' is concerned.

\section{QUANTUM CLONING MACHINES}

Lest one become complacent in accepting the ``obviousness'' of 
$\theta=45^\circ$ signifying when two states are the most quantum 
with respect to each other, let us consider one more notion
of quantumness.  Lately it has become a popular pastime to consider
the issue of how close one can come to ideal cloning for an unknown
quantum state.\refnote{\cite{Deluge}} \ In some ways this is closely 
related to the information-disturbance question; for if one could 
clone ideally, 
then one could create the potential for gathering information without
disturbing.  However, upon closer inspection, one finds quite a
divergence between the two issues.

Consider the issue at hand.  One would like to take the given system,
secretly prepared in either $|\psi_0\rangle$ or $|\psi_1\rangle$, 
attach it to some ancillary system prepared in a standard state 
$|s\rangle\in {\cal H}_A$, and have the two together evolve to the 
state $|\psi_0\rangle|\psi_0\rangle$ or 
$|\psi_1\rangle|\psi_1\rangle$
as should be the case.  Instead, the best one can hope for is some
states $|\psi_0^{\rm\scriptscriptstyle SA}\rangle$ 
and $|\psi_1^{\rm\scriptscriptstyle SA}\rangle$ on the combined space
${\cal H}_S\otimes{\cal H}_A$ that obtain some {\it but
not all\/} the characteristics of ideal clones.  The only
characteristic we shall require here is that the marginal states
isolated to each system be identical, i.e., 
${\rm tr}_{\rm\scriptscriptstyle S}
|\psi_i^{\rm\scriptscriptstyle SA}\rangle
\langle\psi_i^{\rm\scriptscriptstyle SA}|=
{\rm tr}_{\rm\scriptscriptstyle A}|\psi_i^{\rm\scriptscriptstyle SA}
\rangle\langle\psi_i^{\rm\scriptscriptstyle SA}|$ for $i=0$ and 1.

With this as the sole criterion of a cloning device, what is a good 
clone if it is not ideal?  There are several measures that one might
imagine for gauging this, but we shall consider only two: the average
global fidelity $|\langle\psi_i^{\rm\scriptscriptstyle SA}|
(|\psi_i\rangle|\psi_i\rangle)|^2$, and the average local fidelity
$\langle\psi_i|\Big({\rm tr}_{\rm\scriptscriptstyle A}
|\psi_i^{\rm\scriptscriptstyle SA}\rangle
\langle\psi_i^{\rm\scriptscriptstyle SA}|\Big)|\psi_i\rangle$.
Remarkably, these two measures are not optimized by the
same cloning interaction; they give distinct
notions of optimal cloning.\refnote{\cite{Bruss98}}  For the case
at hand, the optimal values of the global and local fidelities turn
out to be
\be
F_g=\frac{1}{2}\Big(1+x^3+(1-x^2)\sqrt{1+x^2}\Big)\;,\qquad\mbox{and}
\ee
$$
F_l=\frac{1}{2}+\frac{\sqrt{2}}{32x}(1+x)\Big(3-3x+\sqrt{1-2x+9x^2}
\Big)\sqrt{-1+2x+3x^2+(1-x)\sqrt{1-2x+9x^2}}\;,
$$
respectively.  Each of these measures now define a notion of
quantumness for our two states.  With respect to $F_g$, two states
are the most quantum with respect to each other when $x=1/\sqrt{3}$.
With respect to $F_l$, they are the most quantum when $x=1/2$.  In 
neither case do we find the coveted $x=1/\sqrt{2}$ value.

\section{CONCLUSION}

What is the essence of quantum theory?  What crucial features of the
phenomena about us lead ineluctably to just this formalism?
These are questions that have been asked since the earliest days of
the theory.  Each generation has its answer; ours no doubt will
find part of it written in the language of quantum information. 
What is striking about the newest turn---the quantum information 
revolution---is that it provides a set of tools for this analysis
from {\it within\/} quantum theory.  The example of the 
tradeoff between information and disturbance in quantum eavesdropping
is typical.  Words about ``measurements causing disturbance'' have
been with us since 1927, but those always in reference to outdated,
illegitimate classical concepts.  The time is ripe to 
consider turning the tables, to ask ``What is quantum mechanics 
trying to tell us?''\refnote{\cite{Mermin98}} \
Why is the world so constituted as to allow single-bit information
transfers to be disturbed by outside information-gatherers, but never
{\it necessarily\/} more so than by an amount 
$D_{\rm\scriptscriptstyle @MI}\approx0.067$?  Why is the world so
constituted that binary preparations can be put together in a
way that the whole is more than a sum of the parts, but never more
so than by $Q\approx0.202$ bits?  The answers surely cannot be that
far away.

\section{ACKNOWLEDGMENTS}
This paper is dedicated to Jeff Kimble for joking that ``there just
can't be so much to say about only two nonorthogonal states.''  I 
thank Micha\l\ Horodecki for useful discussions.  I acknowledge a Lee
A. DuBridge Fellowship and the support of DARPA through the Quantum 
Information and Computing Institute (QUIC).



\begin{numbibliography}

\bibitem{Kraus83}%
K.~Kraus, ``States, Effects, and Operations,'' Springer-Verlag, 
Berlin (1983).

\bibitem{Pitowsky98}%
For a very basic result in this respect, see Theorem 6 and Section V
of I.~Pitowsky, Infinite and finite Gleason's theorems and the
logic of indeterminacy, {\it J.\ Math.\ Phys.} 39:218 (1998).

\bibitem{Wootters82}%
W.~K. Wootters and W.~H. Zurek, A single quantum cannot be 
cloned, {\it Nature\/} 299:802 (1982);
D.~Dieks, Communication by EPR devices, {\it Phys.\ Lett.\ A\/} 
92:271 (1982).

\bibitem{Yuen86}%
H.~P. Yuen, Amplification of quantum states and noiseless photon
amplifiers, {\it Phys.\ Lett.\ A\/} 113:405 (1986);
H.~Barnum, C.~M. Caves, C.~A. Fuchs, R.~Jozsa, and B.~Schumacher,
Noncommuting mixed states cannot be broadcast, {\it Phys.\
Rev.\ Lett.} 76:2818 (1996).

\bibitem{Bennett92b}%
C.~H. Bennett, G.~Brassard, and N.~D. Mermin, Quantum cryptography 
without Bell's theorem, {\it Phys.\ Rev.\ Lett.} 68:557 (1992).

\bibitem{Fuchs98}%
C.~A. Fuchs, Information gain vs.\ state disturbance in 
quantum theory, {\it Fort.\ der Phys.} 46:535 (1998);
C.~A. Fuchs and A.~Peres, Quantum state
disturbance vs.\ information gain:\ Uncertainty relations for
quantum information, {\it Phys.\ Rev.\ A\/} 53:2038 (1996).


\bibitem{Sasaki96}%
M.~Sasaki, T.~S. Usuda, O.~Hirota, and A.~S. Holevo, Applications
of the Jaynes-Cummings model for the detection of nonorthogonal
quantum states, {\it Phys.\ Rev.\ A\/} 53:1273 (1996).

\bibitem{Fuchs97}%
C.~A. Fuchs, Nonorthogonal quantum states maximize classical
information capacity, {\it Phys.\ Rev.\ Lett.} 79:1163 (1997).

\bibitem{Helstrom76}%
C.~W. Helstrom, ``Quantum Detection and Estimation
Theory,'' Academic Press, NY (1976).

\bibitem{Caves96}%
C.~M. Caves and C.~A. Fuchs, Quantum information:~How much
information in a state vector?, {\it in}: ``The Dilemma of 
Einstein, Podolsky and Rosen -- 60 Years Later,'' A.~Mann 
and M.~Revzen, eds., Israel Physical Society (1996).

\bibitem{Bennett98}%
C.~H. Bennett, D.~P. DiVincenzo, C.~A. Fuchs, T.~Mor, E.~Rains, P.~W.
Shor, J.~A. Smolin, and W.~K. Wootters, Quantum nonlocality without
entanglement, 
{\tt quant-ph/9804053}.

\bibitem{Cover91}%
T.~M. Cover and J.~A. Thomas, ``Elements of Information Theory,''
Wiley, NY (1991), Sect.\ 8.9.

\bibitem{Holevo79}%
A.~S. Holevo, Capacity of a quantum communication channel,
{\it Prob.\ Info.\ Trans.} 15:247 (1979).

\bibitem{Fujiwara98}%
A.~Fujiwara and H.~Nagaoka, Operational capacity and 
pseudoclassicality of a quantum channel, {\it IEEE Trans.\ Inf.\
Theory\/} 44:1071 (1998).

\bibitem{Levitin95}%
L.~B. Levitin, Optimal quantum measurements for two pure and mixed
states, {\it in}:  ``Quantum Communications and Measurement,'' V.~P. 
Belavkin, O.~Hirota, and R.~L. Hudson, eds., Plenum Press, NY (1995);
C.~A. Fuchs and C.~M. Caves, Ensemble-dependent bounds for
accessible information in quantum mechanics, {\it Phys.\ Rev.\
Lett.} 73:3047 (1994).

\bibitem{Hausladen96}%
P.~Hausladen, R.~Jozsa, B.~Schumacher, M.~Westmoreland, and
W.~K. Wootters, Classical information capacity of a quantum 
channel, {\it Phys.\ Rev.\ A\/} 54:1869 (1996);
A.~S. Holevo, The capacity of the quantum channel with general
signal states, {\it IEEE Trans.\ Inf.\ Theory\/} 44:269 (1998);
B.~Schumacher and M.~D. Westmoreland,  Sending classical
information via noisy quantum channels, {\it Phys.\ Rev.\ A\/}
56:131 (1997).

\bibitem{Fuchs96a}%
C.~A. Fuchs, ``Distinguishability and Accessible Information in
Quantum Theory,'' Ph.D. thesis, University of New Mexico, 1996.
LANL archive {\tt quant-ph/9601020}.

\bibitem{Peres91}%
A.~Peres and W.~K. Wootters, Optimal detection of quantum 
information, {\it Phys.\ Rev.\ Lett.} 66:1119 (1991).

\bibitem{Pauli95}%
W.~Pauli, ``Writings on Philosophy and Physics,'' 
Springer-Verlag, Berlin (1995).

\bibitem{Peres93}%
A.~Peres, ``Quantum Theory: Concepts and Methods,'' Kluwer, 
Dordrecht (1993).

\bibitem{Bennett92a}%
C.~H. Bennett, Quantum cryptography using any two 
nonorthogonal states, {\it Phys.\ Rev.\ Lett.} 68:3121 (1992).

\bibitem{Deluge}%
For a deluge of articles on cloning, see the LANL {\tt quant-ph} 
archive.

\bibitem{Bruss98}%
D.~Bru\ss, D.~P. DiVincenzo, A.~Ekert, C.~A. Fuchs, C.~Macchiavello,
and J.~A. Smolin, Optimal universal and state-dependent quantum
cloning, {\it Phys.\ Rev.\ A\/} 57:2368 (1998).

\bibitem{Mermin98}%
N.~D. Mermin, What is quantum mechanics trying to tell us?,
{\it Am.\ J. Phys.} 66:753 (1998).

\end{numbibliography}
\end{document}